\documentstyle[12pt]{article}
\begin{document}
\renewcommand{\theequation}{\thesection.\arabic{equation}}
\begin{titlepage}
        \title{Higher--Order Corrections to \\
        the Effective Gravitational Action from Noether Symmetry Approach}
\author{S. Capozziello\thanks{E-mail: capozziello@vaxsa.csied.unisa.it,
capozziello@na.infn.it}
and G. Lambiase\thanks{E-mail:
lambiase@physics.unisa.it} \\
 {\em Dipartimento di Scienze Fisiche "E.R. Caianiello"} \\
 {\em Istituto Nazionale di Fisica Nucleare, Sez. di Napoli,} \\
 {\em Universit\'a di Salerno, 84081 Baronissi (Sa), Italy.} \\ }
\date{\today}
\maketitle
\begin{abstract}

Higher--order corrections of Einstein--Hilbert action of general
relativity can be recovered by imposing the existence of a Noether
symmetry to a class of theories of gravity where Ricci scalar $R$
and its d'Alembertian $\Box R$ are present. In several cases, it
is possible to get exact cosmological solutions or, at least, to
simplify dynamics by recovering  constants of motion. The main
result is that a Noether vector seems to rule the presence of
higher--order corrections of gravitaty.

\end{abstract}
\thispagestyle{empty} \vspace{20.mm}
 PACS number(s): 98.80 H, 0450 \\

\vspace{5.mm}

\vfill

\end{titlepage}

\section{\normalsize\bf Introduction}
\setcounter{equation}{0}

The issue of recovering a suitable effective action seems to be one
of the main ways to construct a consistent theory of quantum
gravity \cite{odintsov}. Starting from pioneering works of Sakharov
\cite{sakharov}, the effects of vacuum polarization on the
gravitational constant, i.e. the fact that gravitational constant
can be induced by vacuum polarization, have been extensively
investigated. All these attempts led to take into account
gravitational actions extended beyond the simple Einstein--Hilbert
action of general relativity which is linear in the Ricci scalar
$R$.

At the beginning, the motivation was to investigate alternative
theories in order to see if gravitational effects could be
recovered in any case.

The Brans--Dicke approach is one of this attempt which, asking for
dynamically inducing the gravitational coupling by a scalar field,
is more coherent with the Mach principle requests \cite{brans}.

Besides, it has been realized that corrective terms are
inescapable if we want to obtain the effective action of quantum
gravity on scales close to the Planck length (see e.g. \cite{vilkovisky}).
In other words, it seems that, in order to construct a
renormalizable theory of gravity, we need higher--order terms of
curvature invariants such as $R^2, R^{\mu\nu}R_{\mu\nu},
R^{\mu\nu\alpha\beta}R_{\mu\nu\alpha\beta}, R\Box R$,
$R\Box^kR$ or nonminimally coupled terms between scalar fields
 and geometry as $\varphi^2R$.

Stelle \cite{stelle} constructed a renormalizable theory of
gravity by introducing quadratic terms in curvature invariants.
Barth and Christensen gave a detailed analysis of the one--loop divergences
of fourth--order gravity theories giving the first general scheme of
quantization of higher--order theories \cite{christensen,barth}.
Several results followed  and today it is well
known that a renormalizable theory of gravity is obtained, at
one--loop level, if at least quadratic terms in the Riemann
curvature tensor and its contractions are introduced
\cite{odintsov}. Any action, where a finite number of terms
involving power laws of curvature tensor or its derivatives appears, is a
low--energy approximation to some fundamental theory of gravity
which, up to now, is not available. For example, string theory or
supergravity present low--energy effective actions where
higher--order or nonminimally coupled terms appear.
\cite{fradkin}.

However, if Lagrangians with higher--order terms or arbitrary
derivatives in curvature invariants are considered, they are
expected to be non--local and give rise to some characteristic
length $l_0$ of the order of Planck length. The expansion in terms
of $R$ and $\Box R$, for example, at scales larger than $l_0$
produces infinite series which should break near $l_0$
\cite{birrel}.

With these facts in mind, taking into account such Lagrangians,
means to make further steps toward a complete renormalizable
theory of gravity. For instance, Vilkovisky \cite{vilkovisky}
considered a non--local Lagrangian of the form $R f(\Box )R$,
where
 \begin{equation}\label{eq1}
 f(\Box)=\int\frac{\rho(x)}{\Box-x}dx\,{,}
 \end{equation}
 in order to construct an effective action of quantum gravity.
Immediately, one realizes that it can be approximated by the sum
 \begin{equation}\label{eq2}
 \sum_{i=0}^{k}R\Box^i R\,{,}
 \end{equation}
 so that we get $(2k+4)$--order field equations.
Also the case ${\displaystyle R\frac{1}{\Box}R}$ has to be mentioned
since it can be
regarded as a conformal anomaly \cite{martin}.

We have to do an
important remark at this point. In the effective action, terms as
$R^j\Box^iR$ have to be taken into account since pure terms as
$\Box^iR$ are total divergences and can be ignored. Terms as
$\Box^iR\Box^jR$ can be integrated by parts giving $R\Box^{i+j}R$
\cite{wands,kluske,schmidt1}.

We are restricting to Lagrangians containing Ricci scalar and its
derivative since in this case it is quite straightforward to
obtain conformal transformations relating higher--order gravity
theories to general relativity with a certain number of scalar
fields \cite{gottloeber}.

Furthermore, only these Lagrangians are interesting for
constructing cosmological models (at least if we restrict the
discussion to homogeneous cosmological models).

However, we have to consider the fact that in the limit of the
classical theory, higher--order Lagrangians give rise to
superfluous degrees of freedom. This is a controversy in
literature \cite{sixth,battaglia,schmidt2}, which, in our
knowledge is not solved. Some authors \cite{parker}, discuss the
possibility that only solutions which are expandable in powers of
$\hbar$ are {\it self--consistent}, others \cite{liddle} consider
such superfluous degrees of freedom as phases of oscillations
around the Friedman behaviour \cite{mueller}. In any case, the
transition to our classical observed universe has to be accurately
discussed.

In this paper, we are going to discuss if such extra--terms in
the effective gravitational action can be recovered
by asking for  symmetries
of a Lagrangian which generic form is
 \begin{equation}\label{eq3}
 {\cal L}=\sqrt{-g}\,F(R, \Box R)\,{.}
 \end{equation}
 We are using the so called {\it Noether Symmetry Approach} which
was extensively used to study nonminimally coupled theories of the
form
 \begin{equation}\label{eq4}
 {\cal L}=\sqrt{-g}\,\left[F(\varphi)R+\frac{1}{2}\nabla_{\mu}\varphi
 \nabla^{\mu}\varphi-V(\varphi)\right]\,{,}
 \end{equation}
 where $\nabla_{\mu}$ is the covariant derivative.
In \cite{pla,cqg4,marek,india}, it was shown that asking for the existence of a
Noether symmetry
 \begin{equation}\label{eq5}
  L_{X}{\cal L}=0\to X{\cal L}=0\,{,}
 \end{equation}
 where $ L_{X}$ is the Lie derivative with respect to the Noether vector
$X$, it is possible to select physically interesting forms of the
interaction potential $V(\varphi)$ and the gravitational coupling
$F(\varphi)$. The scalar field $\varphi$ is generic and it can
represent the matter counterpart in an early universe dynamics.

The existence of Noether symmetries allows to select constants of
motion so that the dynamics results simplified. Often
such a dynamics is exactly solvable by a straightforward change of
variables where a cyclic one is present.

Here we want to apply the same method to higher--order theories of
the form (\ref{eq3}). In particular, we take into account
suitable minisuperspaces  whose
degrees of freedom are $a$, the scalar factor of the universe, $R$,
the Ricci scalar, and $\Box R$ the d'Alembertian of Ricci scalar,
related among them by some Lagrange constraints.
As we shall see in next section, using a
Friedman--Robertson--Walker (FRW) metric it is possible to reduce
the Lagrangian (\ref{eq3}) to a point--like one and then apply the
Noether technique. The main result is that several
fourth, sixth and eight--order interesting
Lagrangians are recovered by asking for the Noether
symmetry. For example, a term as $ \sqrt{R\Box R}$, which is a
part of the so--called $a_3$ anomaly \cite{zeldovich,gilkey}
is connected to the existence of the Noether symmetry.

A similar result works for $F_0R^{3/2}$ which is related to the
Liouville field theory by a conformal transformation. The same
technique gives Lagrangians like
$$ {\cal L}=\sqrt{-g}\,(F_1R+F_2R^2)\,,\quad
\mbox{ or}\quad
{\cal L}=\sqrt{-g}\,(F_1R+F_2R^2+F_3R\Box R)\,,$$
where $F_{i}$ are constants,
 widely studied in  literature, e.g.
 \cite{schmidt2,mueller,starobinski,mijic,maeda,berkin,ferraris}.
 In conclusion, it seems that the Noether approach is related to the
recovering of one--loop and trace anomaly corrections of quantum
gravity.

The paper is organized as follows. Sect.2 is devoted to the
discussion of the generic point--like Lagrangian which can be
recovered from theories like (\ref{eq3}). In Sect.3, we apply
Noether vector to such a Lagrangian showing that the existence of
a symmetry selects its form. Fourth--order and higher than fourth--order
models are discussed in Sect.4 and 5, respectively. In both
cases we study the related FRW cosmology.
 Discussion and conclusions are drawn in Sect.6.

\section{\normalsize\bf Higher-Order Point-like Lagrangians
and Equations of Motion}
\setcounter{equation}{0}

A generic higher--order theory in four dimensions can be described
by the action
 \begin{equation}\label{eq6}
 {\cal A}=\int d^4x \sqrt{-g} F(R, \Box R, \Box^2R, \ldots, \Box^kR)\,{.}
 \end{equation}
We are using physical units $8\pi G_N=c=\hbar=1$. Equations of
motion can be deduced by the method worked out in
\cite{schmidt1,buchdahl}
 \begin{eqnarray}
 G^{\mu\nu}&=&\frac{1}{{\cal G}}\left\{\frac{1}{2}g^{\mu\nu}(F-{{\cal G}}R)
+(g^{\mu\lambda}
 g^{\nu\sigma}-g^{\mu\nu}g^{\lambda\sigma}){\cal G}_{;\,\lambda\sigma}+\right.
 \nonumber \\
    & + &
    \frac{1}{2}\sum_{i=1}^k\sum_{j=1}^i(g^{\mu\nu}g^{\lambda\sigma}+g^{\mu\lambda}
 g^{\nu\sigma})(\Box^{j-i})_{;\,\sigma}\left(\Box^{i-j}\frac{\partial F}{\partial\Box^iR}
 \right)_{;\,\lambda}+\nonumber \\
   &
   -&\left. g^{\mu\nu}g^{\lambda\sigma}\left[(\Box^{j-1})_{;\,\sigma}\Box^{i-j}
   \frac{\partial F}{\partial\Box^iR}\right]\right\}\,{,}
   \label{eq7}
   \end{eqnarray}
   where
 \begin{equation}\label{eq8}
 G^{\mu\nu}=R^{\mu\nu}-\frac{1}{2}g^{\mu\nu}R\,{,}
 \end{equation}
 is the Einstein tensor, and
 \begin{equation}\label{eq9}
 {\cal G}=\sum_{j=0}^{k}\Box^j \left(\frac{\partial
 F}{\partial\Box^jR}\right)\,{.}
 \end{equation}
 As we said, these are pure gravity $(2k+4)$--order field
equations. Matter can be taken into account by introducing the
stress--energy tensor $T^{\mu\nu}$ of a (non)minimally coupled
scalar field \cite{wands,lambda}.

For the sake of simplicity, let us restrict to the Lagrangian (\ref{eq3}).
In this case, we have eight--order field
equations which becomes of sixth--order if the theories is linear
in $\Box R$. To apply Noether Symmetry Approach, let us take into
account the point--like FRW Lagrangian
 \begin{equation}\label{eq10}
{\cal L}={\cal L}(a, \dot{a}, R, \dot{R}, \Box R, \dot{(\Box
R)})\,{.}
 \end{equation}
The variables $R$ and $\Box R$ can be considered
independent  and, by the method of Lagrange multipliers,
we can eliminate higher than one time derivatives  (for fourth--order
case see e.g. \cite{vilenkin}). The action related to Lagrangian
(\ref{eq3}) becomes
 \begin{equation}\label{eq11}
 {\cal A}=2\pi^2\int dt\left\{
a^3F-\lambda_1\left[R+6\left(\frac{\ddot{a}}{a}+\left(\frac{\dot{a}}{a}\right)^2
+\frac{k}{a^2}\right)\right]-\lambda_2\left[\Box
R-\ddot{R}-3\,\left(\frac{\dot{a}}{a}\right)\dot{R}\right]\right\}\,{.}
 \end{equation}
 $\lambda_{1,2}$ are given by varying the action with respect
to $R$ and $\Box R$, that is
 \begin{equation}\label{eq12}
\lambda_1=a^3\,\frac{\partial F}{\partial R}\,{,} \quad
\lambda_2=a^3\frac{\partial F}{\partial (\Box R)}\,{.}
 \end{equation}
 After an integration by parts, the (Helmholtz type) point--like
Lagrangian is
 \begin{equation}
 {\cal L}=6a\dot{a}^2\frac{\partial F}{\partial R}+6a^2\dot{a}\frac{d}{dt}
 \,\left(\frac{\partial F}{\partial R}\right) -a^3\dot{R}\frac{d}{dt}
 \left(\frac{\partial F}{\partial (\Box R)}\right)
 + a^3\left[F-\left(R+\frac{6k}{a^2}\right)\frac{\partial F}{\partial
R}-\Box R\,\frac{\partial F}{\partial (\Box R)}\right]
 \,{.}\label{eq13}
 \end{equation}
A remark is necessary at this point. We can also take into account
 \begin{equation}\label{eq14}
 \lambda_1=a^3\left[\frac{\partial F}{\partial R}+\Box
 \frac{\partial F}{\partial (\Box R)}\right]\,{,}
 \end{equation}
 as a Lagrange multiplier \cite{lambda}.
The Lagrangian which comes out
differs from (\ref{eq13}) just for a term vanishing on the
constraint, being
\begin{equation}
 \tilde{\cal L}={\cal L}-
 a^3\left\{R+6\left[\frac{\ddot{a}}{a}+\left(\frac{\dot{a}}{a}\right)^2
 +\frac{k}{a^2}\right]\right\}\Box\frac{\partial F}{\partial (\Box R)}\,.
\end{equation}
>From this point of view, considering the point--like Lagrangian ${\cal L}$
or $\tilde{\cal L}$ is completely equivalent (this remark is obvious dealing
with the equations of motion).

Let us now derive the Euler--Lagrangian equations from
(\ref{eq13}). They can be also deduced from the Einstein equations
(\ref{eq7}). For the sake of clarity, let us derive them step by step.
The equation
 \begin{equation}\label{eq15}
 \frac{d}{dt}\frac{\partial{\cal L}}{\partial\dot{a}}=
 \frac{\partial{\cal L}}{\partial a}
 \end{equation}
 gives
 \begin{eqnarray}
 & &\left[R\frac{\partial F}{\partial R}+\Box R \frac{\partial
F}{\partial (\Box R)}
 -F\right]+2\left[3H^2+2\dot{H}+
 \frac{k}{a^2}\right]\frac{\partial F}{\partial R}+ \nonumber \\
 &+&2\left[\Box R-H\dot{R}\right]\frac{\partial^2 F}{\partial R^2}
 +\dot{R}\dot{(\Box R)}\frac{\partial^2 F}{\partial (\Box R)^2}+
 \left[2\Box^2R-2H\dot{(\Box R)}+\dot{R}^2\right]
 \frac{\partial^2 F}{\partial R\partial (\Box R)}+ \nonumber \\
 &+&2\dot{R}^2\frac{\partial^3 F}{\partial R^3}+2\dot{(\Box R)}^2
 \frac{\partial^3 F}{\partial R\partial (\Box R)^2}+4\dot{R}\dot{(\Box R)}
 \frac{\partial^3 F}{\partial R^2\partial (\Box R)}=0\,{.}
 \label{eq16}
 \end{eqnarray}
 The equation
 \begin{equation}\label{eq17}
 \frac{d}{dt}\frac{\partial{\cal L}}{\partial\dot{R}}=
 \frac{\partial{\cal L}}{\partial R}
 \end{equation}
gives
 \begin{equation}\label{eq18}
 \Box\frac{\partial{\cal F}}{\partial (\Box R)}=0\,{.}
 \end{equation}
 Finally,
  \begin{equation}\label{eq19}
 \frac{d}{dt}\frac{\partial{\cal L}}{\partial\dot{(\Box R)}}=
 \frac{\partial{\cal L}}{\partial \Box R}\,{,}
 \end{equation}
 coincides with the Lagrange constraints
 \begin{equation}\label{eq20}
 \Box R=\ddot{R}+3H\dot{R}\,{,}
 \end{equation}
 \begin{equation}\label{eq21}
 R=-6\left(\dot{H}+2H^2+\frac{k}{a^2}\right)\,{.}
 \end{equation}
 Here $H=\dot{a}/a$ is the Hubble parameter. The condition on the energy
 \begin{equation}\label{eqen}
 E_{{\cal L}}\equiv \dot{a}\frac{\partial {\cal L}}{\partial\dot{a}}+
 \dot{R}\frac{\partial {\cal L}}{\partial\dot{R}}+
 \dot{(\Box R)}\frac{\partial {\cal L}}{\partial\dot{(\Box R)}}-{\cal
 L}=0\,{,}
 \end{equation}
 which is the $(0,0)$--Einstein equation, gives
 \begin{equation}\label{eq22}
 H^2\left(\frac{\partial F}{\partial R}\right)+
H\frac{d}{dt}\left(\frac{\partial F}{\partial R}\right)
 +\frac{\Gamma}{6}=0\,{,}
 \end{equation}
 where
 \begin{equation}\label{eq23}
 \Gamma=\left(R+\frac{6k}{a^2}\right)\frac{\partial F}{\partial
 R}+\Box R \frac{\partial F}{\partial (\Box R)}-F-\dot{R}\frac{d}{dt}
 \left(\frac{\partial F}{\partial (\Box R)}\right)\,{,}
 \end{equation}
 can be interpreted as a sort of effective density (see also \cite{lambda}).

\section{\normalsize\bf Noether Symmetry Approach}
\setcounter{equation}{0}

A Noether symmetry for the Lagrangian (\ref{eq13}) exists if the
condition (\ref{eq5}) holds. It is nothing else but the
contraction of a Noether vector $X$, defined on the tangent space
$TQ=\{q_i, \dot{q}_i\}$  of
the Lagrangian ${\cal L}={\cal L}(q_{i},\dot{q}_{i})$,
with the Cartan one--form, generically
defined as
 \begin{equation}\label{eq24}
 \theta_{{\cal L}}\equiv \frac{\partial {\cal
 L}}{\partial\dot{q}_i}\,dq^i\,{.}
 \end{equation}
 Condition (\ref{eq5}) gives
 \begin{equation}\label{eq25}
 i_X\theta_{{\cal L}}=\Sigma_0\,{,}
 \end{equation}
 where $i_X$ is the inner derivative and $\Sigma_0$ is the
conserved quantity \cite{pla,cqg4,marek,cqgnoe,bianchi}. In other words,
the existence of the symmetry is connected to the existence of a
vector field
 \begin{equation}\label{eq26}
 X=\alpha^i(q)\frac{\partial}{\partial q^i}+\frac{d\alpha^i(q)}{dt}
 \frac{\partial}{\partial \dot{q}^i}\,{,}
 \end{equation}
 where at least one of the components $\alpha^i(q)$ have to be
different from zero. In our case, the tangent space is
 \begin{equation}\label{eq27}
 TQ=\{a, \dot{a}, R, \dot{R}, \Box R, (\dot{\Box R})\}\,{,}
 \end{equation}
 and  the generator of symmetry  is
 \begin{equation}\label{eq28}
 X=\alpha\frac{\partial}{\partial a}+\beta\frac{\partial}{\partial
 R }+\gamma\frac{\partial}{\partial(\Box R)}+
 \dot{\alpha}\frac{\partial}{\partial\dot{a}}+\dot{\beta}\frac{\partial}{\partial\dot{R}}+
 \dot{\gamma}\frac{\partial}{\partial\dot{(\Box R)}}\,\,{.}
 \end{equation}
The functions  $\alpha, \beta, \gamma$ depend on the variables $a, R, \Box R$.
A Noether symmetry exists if at least one of them is different from
zero.
Their analytic forms can be found by expliciting (\ref{eq5}),
which corresponds to a set of ${\displaystyle 1+\frac{n(n+1)}{2}}$
partial differential
equations given by equating to zero the terms in $\dot{a}^2$,
$\dot{R}^2$, $ \dot{(\Box R)}^2$, $\dot{a}\dot{R}$ and so on. In
our specific case, $n=3$ is the dimension of the
configuration space. We get a system of seven partial differential
equations
 \begin{equation}\label{eq29}
 \frac{\partial F}{\partial R}\left(\alpha+2a\frac{\partial\alpha}{\partial a}\right)
 +a\frac{\partial^2 F}{\partial R^2}\left(\beta+a\frac{\partial\beta}{\partial
a}\right)+a\frac{\partial^2 F}{\partial R\partial(\Box
R)}\left(\gamma+a\frac{\partial\gamma}{\partial a}\right)
 =0\,{,}
 \end{equation}
 \begin{equation}\label{eq30}
 -6\frac{\partial^2 F}{\partial R^2}\frac{\partial\alpha}{\partial
 R}+\frac{\partial^2 F}{\partial R\partial(\Box R)}
 \left(3\alpha+2a\frac{\partial\beta}{\partial R}\right)+
 \beta a\frac{\partial^3 F}{\partial R^2\partial(\Box R)}+
 \gamma a\frac{\partial^3 F}{\partial R\partial(\Box R)^2}+
 a\frac{\partial\gamma}{\partial R}\frac{\partial^2 F}{\partial(\Box R)^2}=0\,{,}
 \end{equation}
 \begin{equation}\label{eq31}
 6\frac{\partial^2 F}{\partial R\partial(\Box R)}\frac{\partial\alpha}{\partial (\Box
 R)}-a\frac{\partial^2 F}{\partial(\Box R)^2}\frac{\partial\beta}{\partial(\Box
 R)}=0\,{,}
 \end{equation}
% \begin{equation}\label{eq32}
 $$
 12\frac{\partial F}{\partial R}\frac{\partial\alpha}{\partial R}+
 6\frac{\partial^2 F}{\partial R^2}\left(2\alpha+a\frac{\partial\alpha}{\partial a}+
 a\frac{\partial\beta}{\partial R}\right)+
 a\frac{\partial^2 F}{\partial R\partial(\Box R)}\left(6\frac{\partial\gamma}{\partial R}
 -2a\frac{\partial\beta}{\partial a}\right)-
 $$
 \begin{equation}\label{eq32}
 -a^2\frac{\partial\gamma}{\partial a}\frac{\partial^2
 F}{\partial(\Box R)^2}+
 6\beta a\frac{\partial^3 F}{\partial R^3}+
 6\gamma a\frac{\partial^3 F}{\partial R^2\partial(\Box R)}=0\,{,}
 \end{equation}
% \begin{equation}\label{eq34}
 $$
 12\frac{\partial F}{\partial R}\frac{\partial\alpha}{\partial (\Box R)}
 +\frac{\partial^2 F}{\partial R\partial(\Box R)}\left(12\alpha+
 6a\frac{\partial\alpha}{\partial a}+6a\frac{\partial\gamma}{\partial (\Box R)}\right)
 +6a\frac{\partial^2 F}{\partial R^2}\frac{\partial\beta}{\partial (\Box
 R)}\,-
% \end{equation}
 $$
 \begin{equation}\label{eq34}
 -a^2\frac{\partial^2 F}{\partial(\Box R)^2}\frac{\partial\beta}{\partial a}+
 6\gamma a\frac{\partial^3 F}{\partial R\partial(\Box R)^2}+
 6\beta a \frac{\partial^3 F}{\partial R^2\partial(\Box R)}=0\,{,}
 \end{equation}
% \begin{equation}\label{eq35}
 $$
 \frac{\partial^2 F}{\partial(\Box R)^2}\left(3\alpha+a\frac{\partial\gamma}{\partial (\Box R)}
 +a\frac{\partial\beta}{\partial R}\right)+
 \frac{\partial^2 F}{\partial R\partial(\Box R)}\left(2a\frac{\partial\beta}{\partial (\Box
 R)}-6\frac{\partial\alpha}{\partial R}\right)
-6\frac{\partial^2 F}{\partial R^2}
 \frac{\partial\alpha}{\partial \Box R}\,+
% \end{equation}
 $$
 \begin{equation}\label{eq35}
 +\beta a\frac{\partial^3 F}{\partial R\partial(\Box R)^2}+
 \gamma a\frac{\partial^3 F}{\partial(\Box R)^3}=0\,{,}
 \end{equation}
% \begin{equation}\label{eq36}
 $$
 3\alpha\left(F-R\frac{\partial F}{\partial R}\right)-\beta a R\frac{\partial^2 F}{\partial R^2}
 -3\alpha\Box R\frac{\partial F}{\partial(\Box R)}-
 \gamma a\Box R\frac{\partial^2 F}{\partial(\Box R)^2}-
 a\frac{\partial^2 F}{\partial R\partial(\Box R)}(\beta\Box
 R+\gamma R)-
% \end{equation}
 $$
 \begin{equation}\label{eq36}
 -\frac{6k}{a^2}\left[\alpha\frac{\partial F}{\partial R}+
 \beta a\frac{\partial^2 F}{\partial R^2}+\gamma a
 \frac{\partial^2 F}{\partial R\partial(\Box R)}\right]=0\,{.}
 \end{equation}
 The system is overdetermined and, if solvable, enables one to
assign $\alpha$, $\beta, \gamma$, and $F(R, \Box R)$. In such a case,
we can always transform the Lagrangian (\ref{eq13}) so that
 \begin{equation}\label{eq38}
 {\cal L}(a, \dot{a}, R, \dot{R}, \Box R, \dot{(\Box R)})\to
 {\cal L}(u, \dot{u}, w, \dot{w}, \dot{z})\,{,}
 \end{equation}
 where $z$ is a cyclic variable and the dynamics is simplified.
This change of variables can be easily obtained by the conditions
 \begin{eqnarray}
 i_Xdz & = & \alpha\frac{\partial z}{\partial a}+\beta\frac{\partial z}{\partial
 R}+\gamma\frac{\partial z}{\partial\Box R}=1 \,{,}\label{eq39} \\
 i_Xdw & = & \alpha\frac{\partial w}{\partial a}+\beta\frac{\partial w}{\partial
 R}+\gamma\frac{\partial w}{\partial\Box R}=0 \,{,}\label{eq40} \\
 i_Xdu & = & \alpha\frac{\partial u}{\partial a}+\beta\frac{\partial u}{\partial
 R}+\gamma\frac{\partial u}{\partial\Box R}=0 \,{,}\label{eq41}
 \end{eqnarray}
 which strictly depend on the form of $\alpha, \beta, \gamma$.
Once we solve the dynamics in the system $\{z, w, u\}$, by the
inverse transformation
 \begin{equation}\label{eq42}
 \{z(t), w(t), u(t)\}\to  \{a(t), R(t), \Box R(t)\}\,{,}
 \end{equation}
 we recover the dynamics in our primitive physical variables. Here
$t$ is the cosmic time.

However, we have to stress that we are considering a constrained
dynamics since the variable $a$, $R$, $\Box R$ are related each
other. In next sections, we show that the existence of the Noether
symmetry, i.e. solving the system (\ref{eq29})--(\ref{eq36}),
gives models of physical interest.

\section{\normalsize\bf Fourth--Order Gravity}
\setcounter{equation}{0}

If the Lagrangian (\ref{eq13}) does not depend on $\Box R$, we are dealing
with fourth--order equations of motion. Furthermore, if
$F(R)=R+2\Lambda$, the standard second--order gravity is recovered, being
$\Lambda$ the cosmological constant.

The configuration space is two-dimensional so that system
(\ref{eq29})--(\ref{eq36}) reduces to
 \begin{eqnarray}
 &&\frac{dF}{dR}\left(\alpha+2a\frac{\partial\alpha}{\partial
 a}\right)+a\left(\frac{d^{2}F}{dR^2}\right)\left(\beta
+a\frac{\partial\beta}{\partial a}\right)=0\,{,}\label{eq43} \\
 &&a^2 \left(\frac{d^2F}{dR^2}\right)\frac{\partial\alpha}{\partial
 R}=0\,{,}\label{eq44} \\
 &&2\left(\frac{dF}{dR}\right)\frac{\partial\alpha}{\partial R}+\frac{d^2F}{dR^2}
 \left(2\alpha+a\frac{\partial\alpha}{\partial a}+a\frac{\partial\beta}{\partial
 R}\right)+a\beta\frac{d^3 F}{d R^3}=0\,{,}\label{eq45} \\
 &&3\alpha\left(F-R\frac{dF}{dR}\right)-a\beta R\frac{d^2F}{dR^2}-
 \frac{6k}{a^2}\left(\alpha\frac{dF}{dR}+a\beta\frac{d^2F}{dR^2}\right)=0\,{,}\label{eq46}
 \end{eqnarray}
 where we need only four equations.
 Immediately we see that a Noether symmetry exists for
$F(R)=R+2\Lambda$. Discarding this trivial case, we get the
solution
 \begin{equation}\label{eq47}
 \alpha=\frac{\beta_0}{a}\,{,}\quad
 \beta=-2\beta_0\frac{R}{a^2}\,{,} \quad F(R)=F_0R^{3/2}\,{,}
 \end{equation}
 for any value of the spatial curvature constant $k=0, \pm 1$;
$\beta_0$ and $F_0$ are constants. This case is interesting in
conformal transformations from Jordan frame to Einstein frame
\cite{cqgconf,magnano} since it is possible to give
explicit form of scalar field potential. In fact, if
 \begin{equation}\label{eq48}
 \tilde{g}_{\alpha\beta}\equiv
 \left(\frac{dF}{dR}\right)\,g_{\alpha\beta}\,{,}\quad
 \varphi=\sqrt{\frac{3}{2}}\ln\left(\frac{dF}{dR}\right)\,{,}
 \end{equation}
 we have the conformal equivalence of the Lagrangians
 \begin{equation}\label{eq49}
 {\cal L}=\sqrt{-g}\,F_0R^{3/2}\longleftrightarrow
 \tilde{{\cal L}}=\sqrt{-\tilde{g}}\left[-\frac{\tilde{R}}{2}+
 \frac{1}{2}\nabla_{\mu}\varphi\nabla^{\mu}\varphi-V_0\exp\left(
 \sqrt{\frac{2}{3}}\varphi\right)\right]\,{,}
 \end{equation}
 in our physical units. This is the so--called Liouville field
theory and it is one of the few cases where a fourth--order
Lagrangian can be expressed, in the Einstein frame,
in terms of elementary functions under
a conformal transformation.
 Using Eqs. (\ref{eq39}) and (\ref{eq40}), we get the new variables
 \begin{equation}\label{eq50}
 w=a^2R\,{,}\quad z=\frac{a^2}{2\beta_0}\,{,}
 \end{equation}
 from which  Lagrangian (\ref{eq13}) (without terms in $\Box
R$) becomes
 \begin{equation}\label{eq51}
 {\cal L}=\frac{9\beta_0}{2}\frac{\dot{z}\dot{w}}{\sqrt{w}}-9k\sqrt{w}
 -\frac{1}{2}\sqrt{w^3}\,{.}
 \end{equation}
 A further change of variable $y=\sqrt{w}$ gives
 \begin{equation}\label{eq52}
 {\cal L}=9\beta_0\dot{z}\dot{y}-9ky-\frac{y^3}{2}\,{,}
 \end{equation}
 where $z$ is  cyclic. The dynamics is given by the equations
 \begin{eqnarray}
 &&\dot{y}=0\,{,}  \\
 &&9\beta_0\ddot{z}+9k+\frac{3}{2}\,y^2=0\,{,}\label{eq54} \\
 &&9\beta_0\dot{z}\dot{y}+9ky+\frac{y^3}{2}=0\,{,}\label{eq55}
 \end{eqnarray}
 whose general solution is
 \begin{eqnarray}\label{eq56}
 y(t)&=&\Sigma_0t+y_0\,{,} \\
 z(t)&=&c_4t^4+c_3t^3+c_2t^2+c_1t+c_0\,{.}\label{eq57}
 \end{eqnarray}
 Going back to physical variables, we get the cosmological solution
 \begin{equation}\label{eq58}
 a(t)=\sqrt{2\beta_0}\,[c_4t^4+c_3t^3+c_2t^2+c_1t+c_0]^{1/2}\,{,}
 \end{equation}
 \begin{equation}\label{eq59}
 R(t)=\frac{(\Sigma_0t+y_0)^2}{2\beta_0[c_4t^4+c_3t^3+c_2t^2+c_1t+c_0]}\,{.}
 \end{equation}
 The constants $c_i$ are combinations of the initial conditions.
Their values determine the type of cosmological evolution. For
example, $c_4\neq 0$ gives a power law inflation while, if the
regime is dominated by the linear term in $c_1$, we get a
radiation--dominated stage.

The system (\ref{eq43})--(\ref{eq46}) admits other solutions.
Another interesting one is given by
 \begin{equation}\label{eq60}
 \alpha=0\,{,}\quad \beta=\frac{\beta_0}{a}\,{,}\quad
 F(R)=F_1R+F_2R^2\,{,}
 \end{equation}
 if the condition
 \begin{equation}\label{eq61}
 R=-\frac{6k}{a^2}
 \end{equation}
 is satisfied. Immediately we get the cosmological solution
  \begin{equation}\label{eq62}
  a(t)=a_0t^{1/2}\,{.}
  \end{equation}
 A similar situation works any time that $d^2F/dR^2\equiv F''\neq
0$. The system (\ref{eq43})--(\ref{eq46}) is solved by
 \begin{equation}\label{eq63}
 \alpha=0\,{,}\quad \beta=\frac{\beta_0}{a F''}\,{.}
 \end{equation}
 Condition (\ref{eq61}) has to be satisfied and the radiative
cosmological solution (\ref{eq62}) is recovered. Particularly
interesting, from the point of view of one--loop corrections of
quantum gravity, are polynomial Lagrangian of the form
 \begin{equation}\label{eq64}
 F(R)=\sum_{j=0}^{N}F_jR^j\,{,} \quad N\geq 0\,{,}
 \end{equation}
 where $F_j$ are constant coefficients whose physical dimension
depends on $j$. Cosmological models coming from such theories have
been widely studied (see e.g. \cite{coa,gott}). However, standard
Einstein coupling is recovered if $F_1=-1/2$.

Results are summarized in Table I.

\section{\normalsize\bf Sixth and Eighth--Order Gravity}
\setcounter{equation}{0}

Considering $\Box R$ as a degree of freedom means to take into
account the whole system (\ref{eq29})--(\ref{eq36}) in order to
find some Noether symmetry.  If $F(R,\Box R)$
depends only linearly on $\Box R$, we have a sixth--order
theory, otherwise we are dealing with eighth--order theories.

A simple sixth--order solution of Noether system
(\ref{eq29})--(\ref{eq36}) is recovered if
 \begin{equation}\label{eq65}
 \alpha=\frac{\alpha_0}{\sqrt{a}}\,{,}\quad \beta\,\,\, \gamma\,\, \mbox
 {any}\,{,}\quad F(R, \Box R)=F_1R+F_2\Box R\,{,}\quad k=0\,{.}
 \end{equation}
 Due to the above considerations on pure divergence
\cite{wands,kluske}, this theory reduces to the Einstein one where
standard cosmological solutions are recovered.

For powers of $\Box R$,
we have  Noether symmetries given by
 \begin{equation}\label{eq66}
 \alpha=0\,{,}\quad \beta=\beta_0\,{,}\quad \gamma=0\,{,}
 \quad F(R, \Box R)=F_1R+F_2(\Box R)^n\,{,}\,\, n\geq 2\,{.}
 \end{equation}
However, the theory can assume different forms integrating by parts
\cite{wands,kluske,schmidt1}.
 The  equations of motion are
 \begin{eqnarray}
& & 4F_1\dot{H}+6F_1H^2+2F_1\frac{k}{a^2}-
 F_2(1-n)(\Box
 R)^n-\left(\frac{\dot{R}}{a}\right)\,\Sigma_0=0\,{,}\label{box1} \\
& & -F_2n(n-1)a^3(\Box R)^{n-2}\dot{(\Box
 R)}=\Sigma_0\,{,}\label{box2} \\
& & \Box R -\ddot{R}-3H\dot{R}=0\,{,}\label{box3} \\
& & 6F_1\left[H^2+\frac{k}{a^2}\right]-F_2(1-n)(\Box
 R)^n+\dot{R}\Sigma_0=0\,{.}\label{box4}
 \end{eqnarray}
 $\Sigma_0$ is the Noether constant and $R$ is the cyclic
variable. The standard Newton coupling is recovered, as usual,
for $F_1=-1/2$. A solution for this system is
 \begin{equation}\label{sol}
 a(t)=a_0t\,{,}
 \end{equation}
for $k=-1$, $\Sigma_0=0$,  and for
arbitrary $n,\, F_2$.
Another solution is
 \begin{equation}\label{sol1}
 a(t)=a_0t^{1/2}\,{,}
 \end{equation}
for $k=0$, $\Sigma_0=0$, $F_{1}=0$  and for
arbitrary $n,\, F_2$.
Finally we get
\begin{equation}
\label{sol2}
a(t)=a_{0}\exp(k_{0}t)\,,
\end{equation}
for $k=0$, $\Sigma_{0}=0$, $F_{1}=0.$

 The radiative cosmological solution (\ref{eq62}) with the
condition (\ref{eq61}) is recovered for the cases
 \begin{equation}\label{eq67}
 \alpha=0 \,{,} \quad \beta=\frac{\beta_0}{a}\,{,} \quad
 \gamma\,\, \mbox{any}\,{,}\quad F(R, \Box R)=F_1R+F_2R^2+F_3\Box
 R\,{,}
 \end{equation}
 and
 \begin{equation}\label{eq68}
 \alpha=0 \,{,} \quad \beta=0\,{,}\quad
 \gamma=\frac{\gamma_0}{a}\,{,}
 \quad F(R, \Box R)=F_1R+F_2R^2+F_3R\Box
 R\,{,}
 \end{equation}
 with $\gamma_0$ constant.
The second one is of physical interest and the related
cosmological models have been widely studied \cite{gottloeber,berkin}.

Another Noether symmetry is recovered for
 \begin{equation}\label{eq69}
 \alpha=0 \,{,} \quad \beta=\beta_0\,{,}\quad \gamma=\beta_0\,\frac{\Box
 R}{R}\,{,} \quad F(R, \Box R)=F_1R+F_2\sqrt{R\Box R}\,{,}
 \end{equation}
or simply for
\begin{equation}
F(R, \Box R)=F_2\sqrt{R\Box R}\,{.}
\end{equation}
 This case deserves a lot of attention since $\sqrt{R\Box R}$ is
exactly a part of $a_3$--anomaly \cite{gottloeber} which can be
recovered from the general analysis of one--loop contributions to
gravitational action \cite{zeldovich,gilkey}. Such a fact is
indicative since it seems that searching for Noether symmetries
could be a relevant method for constructing an effective action of
quantum gravity.
 A straightforward change of variables, given by
(\ref{eq39})--(\ref{eq41}), is
 \begin{equation}\label{eq70}
 z=R\,{,}\quad u=\sqrt{\frac{\Box R}{R}}\,{,} \quad w=a\,{.}
 \end{equation}
Choosing the standard Einstein coupling $F_1=-1/2$ the
 Lagrangian (\ref{eq13}) becomes
 \begin{equation}\label{eq71}
 {\cal L}=3[w\dot{w}^2-kw]-
  F_2\left[3w\dot{w}^2u+3w^2\dot{w}\dot{u}+\frac{w^3\dot{z}\dot{u}}{2u^2}
 -3kwu\right]\,{,}
 \end{equation}
where $z$ (i.e. $R$) is the cyclic variable.
>From (\ref{eq71}), it is immediate to derive the equations of motion.
As above, it is possible to recover the particular solutions
\begin{equation}
\label{solu4}
a(t)=a_{0}t\,,\quad a(t)=a_{0}t^{1/2}\,,\quad a(t)=a_{0}\exp(k_{0}t)
\end{equation}
depending on the
set of parameters $\{\Sigma_0, k, F_2\}$. A phase--space view and
conformal analysis, as that in \cite{berkin}, gives the conditions
for he onset and duration of inflation which, specifically,
depends on the sign and the value of $F_2$ and $\Sigma_{0}$.
This fact restricts the
set of initial conditions capable of furnishing satisfactory
inflationary cosmology as discussed in \cite{gottloeber,berkin}.
 In Table II we report the main results of this section.

\section{\normalsize\bf Discussion and Conclusions}

In this paper, we have used the Noether Symmetry Approach in order to
study higher--order theories of gravity. The existence of a
symmetry selects the form of higher--order Lagrangian as in
nonminimally coupled theories (\ref{eq4}); there this technique allows
to assign the form
of the coupling $F(\varphi)$ and the potential $V(\varphi)$
\cite{pla,cqg4}. Here we discussed theories up to eight--order but
it is clear that the method works also for orders beyond. The scheme
is always the same: $i)$ if a symmetry exists, $ii)$ the form of the
effective Lagrangian is assigned, $iii)$ a suitable change of
variables allows to write the dynamics so that a cyclic coordinate
appears.
The solution of the cosmological problem results simplified
since two first integrals of motion are present (the energy and
the symmetry). However the scheme works if a suitable minisuperspace
has been {\it a priori} defined. Here we used FRW minisuperspaces.

Some final remarks are necessary at this point. First of all,
by this technique, one is capable of selecting higher--order
terms as $R^{3/2}$ or $\sqrt{R\Box R}$ of physical interest since
they can be connected to the one--loop or trace anomaly
corrections of the effective action of quantum gravity. It is
worthwhile to stress that such terms are not perturbatively
introduced but emerge by the request of symmetry. Furthermore, the
system of partial differential equations
(\ref{eq29})--(\ref{eq36}) can have several solutions and their
finding out is just a question of mathematical ability. In this paper,
we have not made an exhaustive list of the possible Noether symmetries in
higher--order theories, but we have only presented some examples in fourth-,
sixth-, and eighth--order models.

A further point which has to be stressed is that
obtaining some higher--derivative terms in the effective Lagrangian of
gravity does not, automatically, makes the theory renormalizable. In other
words, functions of $R$ and $\Box R$
alone do not give renormalizable Lagrangians.
For renormalization in fourth--order gravity, see e.g. \cite{barth}.
Our point of view is  that the existence of a Noether symmetry for the
Lagrangian seems to be connected to the existence of corrective terms
which one needs for renormalization and the approach presented in this paper
seems a scheme which could be generalized.

Finally,
 the existence of a Noether symmetry makes
the analysis of a given cosmology more tractable; however
the existence of such a
symmetry is not, in itself, a sufficient motivation for preferring a particular
theory. The situation, as also shown elsewhere \cite{cqg4,marek,india}, becomes
interesting if the selected theory is physically relevant
{\it per se} and the Noether approach selects just it.

A further step  which the authors are going to
achieve is to apply the Noether
Symmetry Approach to the full field theory without reducing to
particular minisuperspaces as done here.

\newpage

\begin{center}
{\bf Table I}-- Symmetries in Fourth--Order Models
\end{center}

\begin{center}
 \begin{tabular}{||l|l|lr||} \hline
   $F(R)$         &   $\alpha$  &  $\beta$   &\\ \hline

   $R+2\Lambda$   &      0      &  $\beta(a, R)$  & \\ \hline

   $F_0R^{3/2}$   & $\beta_0a^{-1}$ & $-2\beta_0a^{-2}R$  &\\ \hline

   $F_0R+F_1R^2$ &      0       &   $\beta_0a^{-1}$    &\\ \hline

   $\sum_{j=0}^{N}F_jR^j$ & 0  & $\beta_0(aF'')^{-1}$ &\\ \hline
  \end{tabular}
\end{center}

\begin{center}
{\bf Table II}-- Symmetries in Higher than Fourth--Order Models
\end{center}

\begin{center}
\begin{tabular}{||l|l|l|lr||}  \hline
   $F(R, \Box R)$    &   $\alpha$   &   $\beta$  & $\gamma$ & \\ \hline
 $F_1R+F_2(\Box R)^n (n\neq 1)$  & 0 & $\beta_0$ & 0  &\\ \hline
 $F_1R+F_2R^2+F_3\Box R$ & 0  &  $\beta_0a^{-1}$ & $\gamma(a, R, \Box R)$ &\\ \hline
 $F_0R+F_1R^2+F_2R\Box R$ & 0  &   0  &  $\gamma_0a^{-1}$ &\\ \hline
 $F_2\sqrt{R(\Box R)}$ & 0 & $\beta_0$ & $\beta_0R^{-1}\Box R$ &\\ \hline
 $F_1R+F_2\sqrt{R(\Box R)}$ & 0 & $\beta_0$ & $\beta_0R^{-1}\Box R$ &\\ \hline
\end{tabular}
\end{center}

\end{document}